\documentclass[american,aps,pra,reprint, superscriptaddress]{revtex4-1}
\usepackage{amsmath,amscd,amsthm,amssymb}
\usepackage{mathrsfs}
\usepackage{graphicx}
\usepackage{epsf,epstopdf}
\usepackage{caption3}
\usepackage[all]{xy}
\usepackage{tikz}
\usepackage[toc,page]{appendix}

\usepackage[unicode=true,pdfusetitle, bookmarks=true,bookmarksnumbered=false,bookmarksopen=false, breaklinks=false,pdfborder={0 0 0},backref=false,colorlinks=false] {hyperref}
\hypersetup{ colorlinks,linkcolor=myurlcolor,citecolor=myurlcolor,urlcolor=myurlcolor}
\usepackage{graphics,epstopdf,graphicx, amsthm, amsmath, amssymb, times, braket, color, bm}
\usepackage[up]{subfigure}
\usepackage{cleveref}

\usepackage{marvosym} 

\definecolor{myurlcolor}{rgb}{0,0,0.7}

\def\be{\begin{equation}}
\def\ee{\end{equation}}
\def\bea{\begin{eqnarray*}}
\def\eea{\end{eqnarray*}}
\def\ot{\otimes}

\theoremstyle{plain}

\providecommand{\theoremname}{Theorem}

\newcommand{\iinner}[2]{\langle #1 | #2\rangle}
\newcommand{\out}[2]{| #1\rangle\langle #2 |}
\DeclareMathOperator{\trace}{tr}
\newcommand{\ptr}[2]{\trace_{#1}({#2})}
\newcommand{\tr}[1]{\ptr{}{#1}}
\newcommand{\id}{\mathbb{I}}

\newcommand*{\myproofname}{Proof}

\makeatother

\def\cH{\mathcal{H}}\def\cI{\mathcal{I}}

\theoremstyle{definition}

\theoremstyle{remark}

\graphicspath{{figs/}}

\begin{document}

 \author{Sunho Kim}
 \email{kimshanhao@126.com}
 \affiliation{School of Mathematical Sciences, Harbin Engineering University, Harbin 150001, People's Republic of China}

 \author{Longsuo Li}
 \email{lilongsuo@126.com}
 \affiliation{Department of Mathematics, Harbin Institute of Technology, Harbin 150001, People's Republic of China}

 \author{Asutosh Kumar}
 \email{asutoshk.phys@gmail.com}
 \affiliation{P.G. Department of Physics, Gaya College, Magadh University, Rampur, Gaya 823001, India}
 \affiliation{Vaidic and Modern Physics Research Centre, Bhagal Bhim, Bhinmal, Jalore 343029, India}
 \affiliation{Harish-Chandra Research Institute, HBNI, Chhatnag Road, Jhunsi, Allahabad 211019, India}

 \author{Chunhe Xiong}
 \email{xiongchunhe@zju.edu.cn}
 \affiliation{Institute of Computer Science Theory, School of Data and Computer Science, Sun Yat-sen University, Guangzhou 510006,   People's Republic of China}

 \author{Sreetama Das}
 \email{sreetama.das21@gmail.com}
 \affiliation{Faculty of Physics, Arnold Sommerfeld Centre for Theoretical Physics (ASC), Ludwig-Maximilians-University Munich, Theresienstr. 37, 80333 Munich, Germany}

 \author{Ujjwal Sen}
 \email{ujjwal@hri.res.in}
 \affiliation{Harish-Chandra Research Institute, HBNI, Chhatnag Road, Jhunsi, Allahabad 211019, India}

 \author{Arun Kumar Pati}
 \email{akpati@hri.res.in}
 \affiliation{Harish-Chandra Research Institute, HBNI, Chhatnag Road, Jhunsi, Allahabad 211019, India}

 \author{Junde Wu}
 \email{wjd@zju.edu.cn}
 \affiliation{School of Mathematical Sciences, Zhejiang University, Hangzhou 310027, People's Republic of China}

\title{Protocol for unambiguous quantum state discrimination using quantum coherence}
\begin{abstract}
Roa \emph{et al.} showed that quantum state discrimination between two nonorthogonal quantum states does not require quantum entanglement but quantum dissonance only. We find that quantum coherence can also be utilized for unambiguous quantum state discrimination. We present a protocol and
quantify the required coherence for this task. We discuss the
optimal unambiguous quantum state discrimination strategy in some
cases. In particular, our work illustrates an avenue to find the optimal
strategy for discriminating two nonorthogonal quantum states by measuring quantum coherence.

\end{abstract}
\maketitle

\section{Introduction}

A fundamental result in
quantum mechanics is the impossibility to perfectly distinguish two
or more nonorthogonal states.
Quantum state discrimination (QSD) consists in devising strategies
to discriminate nonorthogonal quantum states as accurately as
possible.
QSD has various useful applications in quantum information
processing \cite{Helstrom, Waldherr, Bae}, and it branches out into
two important streams:   minimal-error deterministic quantum state
discrimination (DQSD)  \cite{Helstrom} and unambiguous  quantum
state discrimination (UQSD) \cite{Peres}.
In DQSD, one always has an answer but with a probability of being wrong. On the other hand, in UQSD, one is guaranteed to never be wrong, but there are occasions when one does not have an answer. In UQSD, the task is to minimize the probability of no answer.
Though several strategies exist
to discriminate quantum states in the literature, optimal
strategies of QSD are yet to be figured out in all the cases
\cite{Barnett}.
The study of minimization of error in state discrimination was
pioneered by Helstrom \cite{Helstrom} who provided a lower bound on
the error probability for distinguishing two quantum states. It has
been enriched further by presenting an upper bound of success
probability for distinguishing arbitrary number of quantum states
\cite{Zhang}, and many studies have focused on achieving that bound
\cite{Acin, Takeoka, Cook, Wittmann, Tsujino, Assalini}. In
addition, the protocol for unambiguous discrimination of linearly
independent pure quantum states, assisted by an auxiliary system, is
of fundamental interest \cite{Bergou}. While quantum entanglement
\cite{horodecki4} is regarded as a key resource in quantum
information processing \cite{Nielsen}, other non-classical correlations
such as quantum discord and quantum dissonance \cite{ModiPRL, ModiRMP,
Bera2}) are also very useful.
The assisted unambiguous discrimination for two nonorthogonal states that requires only
quantum dissonance (zero entanglement  and nonzero discord)
was introduced by Roa \emph{et al.} \cite{Roa},
and its generalization and various applications have been studied
thereafter \cite{Li, Pang}. An optical implementation of unambiguous
discrimination of the two finite ensembles of coherent states was
also proposed by Sedl\'{a}k \cite{Sedlak}. In this paper, we find
a UQSD protocol that requires only quantum coherence as a resource.


Although those have intrinsically the same origin, viz. the superposition principle, more attention has been
paid on the effects of entanglement and other quantum correlations than on
the impact of quantum coherence \cite{Baumgratz, Aberg} on quantum advantages in devices
and protocols.
The fact that quantum correlations such as entanglement
and dissonance are required to discriminate quantum states, a
natural question arises: is coherence sufficient for
UQSD and is there any relation between the degree
of coherence and the efficiency of discrimination?

In this paper, we answer these questions affirmatively.
In particular, we design a method to find the optimal
UQSD by controlling the coherence in a protocol that discriminates
two nonorthogonal quantum states.
In line with this, we compute the amount of coherence for the optimal UQSD
and determine whether this optimality is achieved by the generated coherence in some circumstances.

In our study, we consider a qudit system $S$ that is randomly prepared in one of the $d$
nonorthogonal but linearly independent pure quantum states. The system $S$ is coupled to
a $(d+1)$-dimensional auxiliary system $A$ by a joint unitary operator $U_{SA}$.
We give a protocol to construct the $U_{SA}$ for $d \ge 2$.
We find that the quantum states post the joint unitary operation do not contain any quantum correlation such as entanglement or quantum discord between the system $S$ and the auxiliary system $A$.
However, quantum coherence is always generated in the auxiliary system $A$ except when the quantum states to be discriminated are mutually orthogonal. The joint unitary thus converts nonorthogonality on the original system $S$ into coherence on the auxiliary system $A$, and this coherence can be consumed for the discrimination of nonorthogonal states.



\section{UQSD with coherence}
%

Quantum coherence \cite{Baumgratz, Aberg} is defined with respect to a fixed orthonormal basis $\{\ket{i}\}$ of a system represented by a Hilbert space $\cH$.
The set of ``incoherent'' or free states is conceptualized as a set of perfectly distinguishable pure states and their mixtures. Precisely, it is defined by $\cI = \big\{\sigma = \sum_ip_i\out{i}{i} : p_i\geq0, \sum_ip_i = 1 \big\}$.
The ``incoherent'' or free operations keep the free states within the set of free states. Precisely, they are completely positive maps, $\Phi$, given by $\Phi(\sigma)= \sum_kE_k\sigma E_k^\dagger$, for a set of incoherent Kraus operators, $\{E_k\}$, so that $\Phi(\sigma) \subseteq \cI$ for all $\sigma \subseteq \cI$.
A measure of coherence (with respect to the von Neumann measurement $\Pi = \{\Pi_i = \out{i}{i}\})$, $C(\rho|\Pi)$, satisfies

(C1) $C(\rho|\Pi) \geq 0$ with equality if and only if $\rho\in \cI$,

(C2) $C(\rho|\Pi)$ is nonincreasing under incoherent operations, i.e., $C(\rho|\Pi)\geq C(\Phi(\rho)|\Pi)$ with $\Phi(\cI) \subseteq \cI$,

(C3) $C(\rho|\Pi)$ is convex in $\rho$.

There are many important coherence measures \cite{Baumgratz, Girolami, Lostaglio, Bera, Shao, Pires, Xi, Rana, Rastegin, Napoli, Luo4, Bu, Yuan, XiongPRA18}. In this paper, we will use two coherence measures. The first coherence measure is an improved version of $K$ coherence \cite{Girolami} based on the Wigner-Yanase skew information $I(\sigma, K) = -\frac{1}{2}\tr{[\sqrt{\sigma}, K]^2}$, proposed by Luo \emph{et al.} and defined as \cite{Luo4}
\be\label{eq:coherence}
C_I(\rho|\Pi) = \sum_i I(\rho, \Pi_i),
\ee
where $I(\sigma, \Pi_i) = -\frac{1}{2}\tr{[\sqrt{\sigma}, \Pi_i]^2}$.

For pure states $\ket{\psi} = \sum_i \psi_i \ket{i}$, this measure is equivalent to the coherence measures such as $l_2$ norm of coherence $C_{l_2}$ and fidelity of coherence $C_f$ \cite{Baumgratz, Shao}:
\bea
C_I(\out{\psi}{\psi}|\Pi) &=& \sum_{i,j,i\neq j}|\psi_i|^2|\psi_j|^2 = C_f(\out{\psi}{\psi}|\Pi)\\
&=& C_{l_2}(\out{\psi}{\psi}|\Pi).
\eea

The second coherence measure $C$ can be either robustness of coherence $C_R$ or $l_1$ norm of coherence $C_{l_1}$ because these measures have the same expression for pure states \cite{Baumgratz, Napoli}.

The axiomatic formulation of the coherence measures paves the way for using any measure without significant digressions in the physics content.
Quantum coherence has been detected experimentally \cite{Girolami, Li2, Zhou, Gao}. Further interesting developments in quantum coherence theory can be explored in Refs. \cite{Hillery, Namit, Yuan, XiongPRA18, Winter, Streltsov, Chitambar2, Marvian, Streltsov2, Streltsov3, Streltsov4, Ma, KimPRA18, XiongPRA19, Theurer}.

In UQSD, one seeks for the best quantum measurement to discriminate between the nonorthogonal states $\ket{\phi_i} \in \cH$ of the ensemble $\{p_i, \ket{\phi_i}\}_{i=1}^d$ with the least possible ``error''. An upper bound on the success probability ($P_s$) of UQSD is given by \cite{Zhang}
\be\label{eq:upper bound success probability}
P_s \leq 1 - \frac{1}{d-1} \sum_{i, j\neq i} \sqrt{p_ip_j}|\iinner{\phi_i}{\phi_j}|.
\ee
This has an operational meaning in the context of duality between the quantum coherence and the path distinguishability \cite{Bera}.\\

Let us consider a qudit that is randomly prepared in one of the $d$
nonorthogonal but linearly independent quantum states $\ket{\phi_i}$
in quantum system $S$, $i=1,2,...,d$ , with probabilities $p_i$. The
system $S$ is coupled to a $(d+1)$-dimensional auxiliary system $A$
by a joint unitary operator $U_{SA}$ such that
\parbox{8.1cm}{
\begin{eqnarray*}
U_{SA}\ket{\phi_i} \ket{0}_A = \sqrt{1-|\alpha_i|^2}\ket{\varphi_i} \ket{i}_A + \alpha_i\ket{\varphi_i} \ket{0}_A,
\end{eqnarray*}}\hfill
\parbox{.3cm}{\begin{eqnarray}\label{eq:joint unitary transformation}\end{eqnarray}}\\
where $\alpha_i^\ast\alpha_j\iinner{\varphi_i}{\varphi_j}=
\iinner{\phi_i}{\phi_j}$ for $i\neq j$. A protocol for constructing the $U_{SA}$ for $d \ge 2$ is discussed in the Appendix A.
After the joint unitary operation $U_{SA}$, the average quantum
state is given as a mixed state $\rho = \sum_{i=1}^d p_i \rho_i =
\sum_{i=1}^d p_i\ket{\varphi_i}\bra{\varphi_i} \ot\rho^A_i,$ where
$\rho_i = U_{SA}(\ket{\phi_i}\bra{\phi_i}
\ot\ket{0}_A\bra{0})U_{SA}^\dagger$ and $\rho^A_i =
(1-|\alpha_i|^2)\ket{i}_A\bra{i} + |\alpha_i|^2\ket{0}_A\bra{0} +
\sqrt{1-|\alpha_i|^2}\big(\alpha_i\ket{0}_A\bra{i} +
\alpha_i^\ast\ket{i}_A\bra{0}\big).$
Note that $\rho^A_i$ is pure for each $i$. If we perform the local measurement
$M = \{\ket{j}_A\bra{j}\}_{j=0}^d$ on the auxiliary system, the success probability
to discriminate the state is given by
\begin{eqnarray}
\label{eq:success probability}
P_s &=& 1- \tr{\id\ot\ket{0}_A\bra{0}\rho} = \sum_{i=1}^d p_i(1-|\alpha_i|^2),
\end{eqnarray}
where $\id$ is the unit operator  for the system $S$.
Also, since $\rho_i^A$ are pure for all $i$, the quantum states post the
unitary operation do not contain any quantum correlation such as
entanglement or quantum discord between the system $S$ and the
auxiliary system $A$. This process only generates and consumes
quantum coherence in the auxiliary system $A$.

Now, we compute the mean of coherence in the basis
$\{\ket{j}_A\}_{i=0}^d$ of the auxiliary system using the measure of coherence defined in equation (1)
with the measurement $\Pi^A = \{\Pi_j^A = \ket{j}_A\bra{j}\}$.
We define the mean of coherence as $C_{mean} := \sum_i p_i
C_I(\rho^A_i|\Pi^A) = \sum_{i=1}^d p_i \Big[\sum_{j=0}^d I(\rho^A_i,
\Pi_j^A)\Big]$ which reduces to
\begin{eqnarray}\label{eq:mean value}
C_{mean}  &=& 2\sum_{i=1}^d p_i |\alpha_i|^2 \big(1 - |\alpha_i|^2\big),
\end{eqnarray}
and $\tilde{C}_{mean} := \sum_{i=1}^d p_iC(\rho^A_i|\Pi^A)$ which reduces to
\begin{eqnarray}\label{eq:mean2}
\tilde{C}_{mean} := 2\sum_{i=1}^d p_i |\alpha_i| \sqrt{1 - |\alpha_i|^2},
\end{eqnarray}
where $C$ can be either robustness of coherence $C_R$ or $l_1$ norm of coherence $C_{l_1}$.
%
%

\begin{figure}[t]
\centering
\includegraphics[height=.24\textheight]{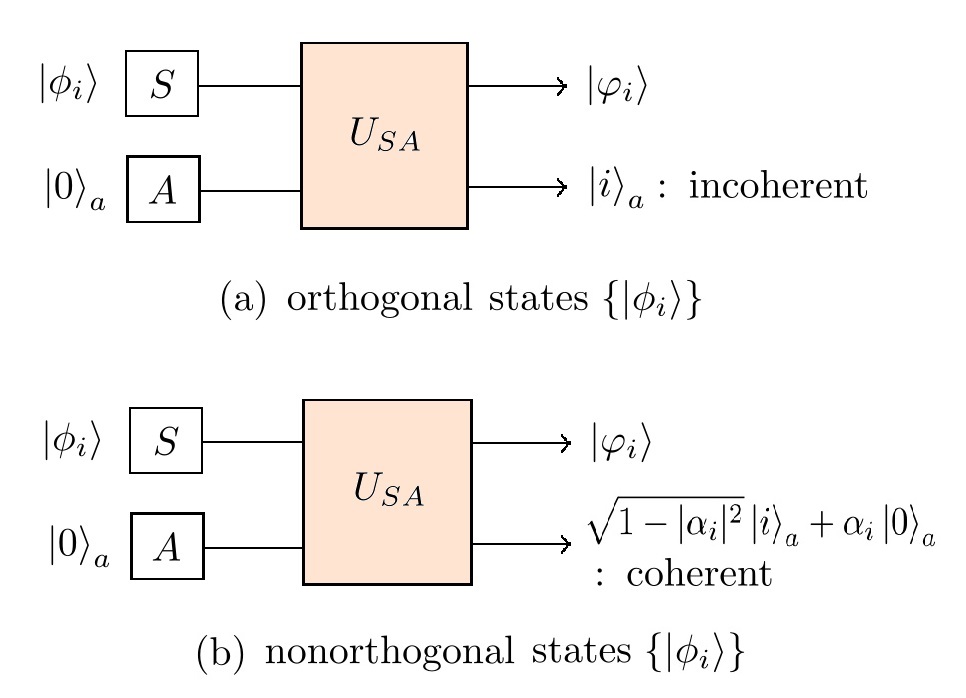}
\caption{\label{fig:1} (a) The UQSD strategy for orthogonal quantum states does not require any coherence. (b) On the contrary, coherence is essential for the UQSD strategy in the case of nonorthogonal quantum states. The degree of nonorthogonality
between the quantum states is closely related to the degree of the generated coherence.}
\end{figure}

This shows that the success probability is lower bounded by the quantum coherence generated in the auxiliary system, i.e., we have
$P_s \ge \frac{1}{2} C_{mean}.$
Another important observation here is that coherence is always
generated except when the quantum states to be discriminated are
mutually orthogonal (see Fig.\ref{fig:1}).

The joint unitary thus converts nonorthogonality
on the original system $S$ into coherence on the auxiliary system $A$, and this coherence
can be consumed for the discrimination of nonorthogonal states (see Ref.
\cite{note2}).

\begin{figure}[htb]
\centering
\includegraphics[height=.13\textheight]{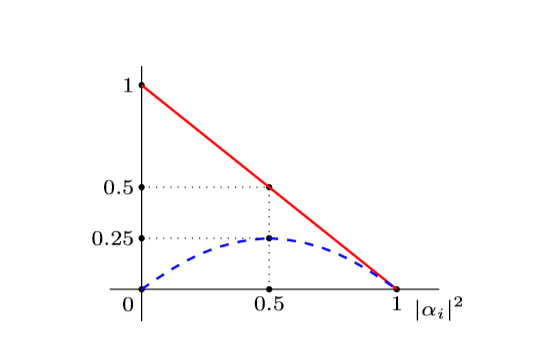}
\caption{\label{fig:fig-2} (Color online) The solid red line is the graph of $1- |\alpha_i|^2$ and the blue dashed line is the graph of $|\alpha_i|^2\big(1- |\alpha_i|^2\big)$.}
\end{figure}

Also from the point of view of each $i$, not the mean of coherence, Eqs. (\ref{eq:mean value}) and (\ref{eq:success probability}) provide us with a heretical relationship between the probability of success $1-|\alpha_i|^2$ and the generated coherence $|\alpha_i|^2\big(1- |\alpha_i|^2\big)$ for each $i$ (see Fig. \ref{fig:fig-2}).

Let us assume that the quantum states $\{\ket{\phi_i}\}_{i=1}^d$
satisfy the condition
$|\iinner{\phi_i}{\phi_j}|\geq\frac{1}{\sqrt{2}}$ for all $i\neq j$,
then we have $|\alpha_i|^2\geq \frac{1}{2}$ for all $i$, because
$|\alpha_i|^2|\alpha_j|^2\geq
|\alpha_i|^2|\alpha_j|^2|\iinner{\varphi_i}{\varphi_j}|^2 =
|\iinner{\phi_i}{\phi_j}|^2\geq\frac{1}{2}.$
In this case, we see from Fig. \ref{fig:fig-2} that $1- |\alpha_i|^2$ decreases when $|\alpha_i|^2\big(1- |\alpha_i|^2\big)$ decreases. This means that if the coherence of $i$-th quantum state after the joint unitary operation is decreased, then the success probability for result $i$ is also decreased.
Conversely, if $|\iinner{\phi_i}{\phi_j}|$ is small enough for all
$i\neq j$ and $|\alpha_i|^2$ is not greater than $\frac{1}{2}$, then
we can increase the probability of success for the result $i$ by
adjusting the $i$-th coherence to be sufficiently small, as seen in
Fig. \ref{fig:fig-2}. However, this is possible only with
independent relationship for each result $i$, and it is difficult to
find a numerical relationship with the optimal UQSD average above.

\section{Mean of coherence for optimal unambiguous discriminations}

Here we consider in detail the two-dimensional case. Recall the UQSD protocol in Eq. (\ref{eq:joint unitary
transformation}) for $d=2$. Because it is always possible to make the nonorthogonal quantum states $\ket{\varphi_1},
\ket{\varphi_2}$ in Eq. (\ref{eq:joint unitary transformation}) the
same (see Appendix A), we have $\alpha_1^\ast\alpha_2= \iinner{\phi_1}{\phi_2} \equiv \gamma$.

\begin{figure}[t]
\centering
\includegraphics[height=.32\textheight]{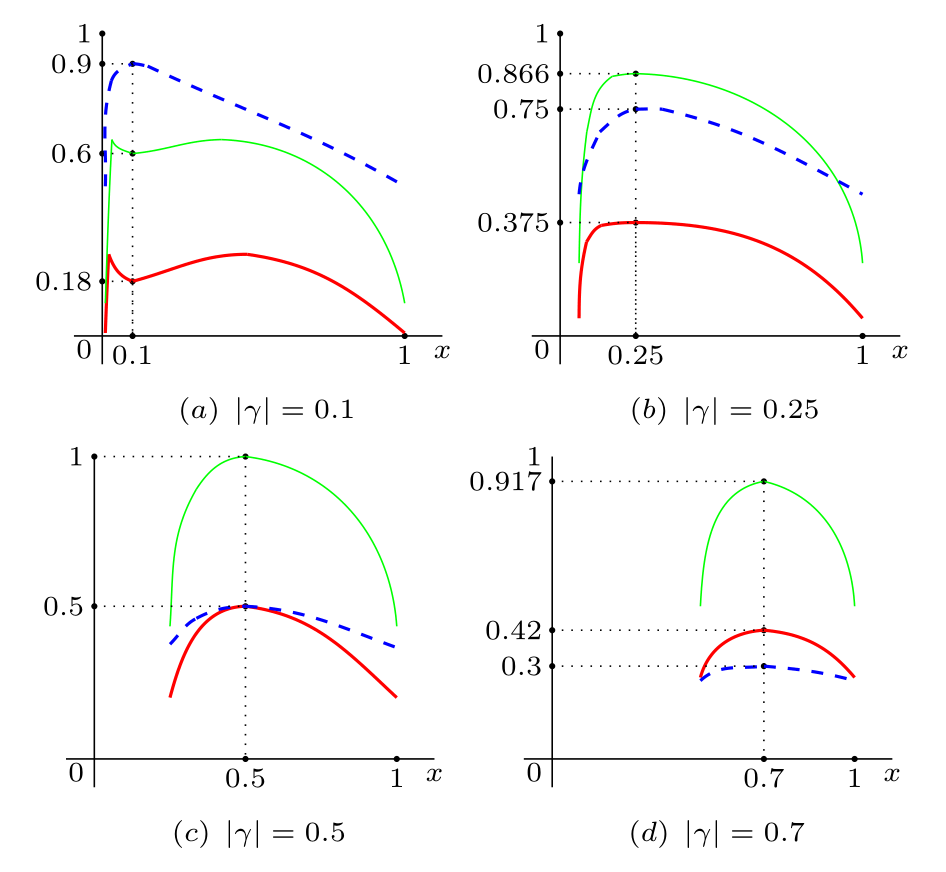}
\caption{\label{fig:fig-3} (Color online) Plots of $P_s|_{x = |\alpha_1|^2}$(dashed blue), $C_{mean}|_{x = |\alpha_1|^2} $(solid red) and $\tilde{C}_{mean}|_{x = |\alpha_1|^2} $ (thin green), when $p_1 = p_2 = \frac{1}{2}$, against $x$ for different values of $|\gamma|$.}
\end{figure}

If $p_1= p_2 = \frac{1}{2}$, we have an interesting
fact that the extreme values of the success probability $P_s$ and
the mean of coherence $C_{mean}$ (or $\tilde{C}_{mean}^{opt}$) are obtained at the same point
$|\alpha_1|^2 = |\alpha_2|^2 = \gamma$; see Appendix
for the detailed calculation. It follows that we
can implement the optimal UQSD strategy by adjusting the mean of
coherence to the maximum value in a defined interval $|\gamma|^2
\leq |\alpha_2|^2 \leq 1$ when $|\gamma|\geq\frac{1}{4}$ (see the
red lines of (b), (c) and (d) in Fig. \ref{fig:fig-3}). Conversely,
when $|\gamma|< \frac{1}{4}$, we can implement the optimal
discrimination by adjusting the mean of coherence to the local
minimum value (see the red line in Fig. \ref{fig:fig-3}(a)).
Furthermore, the same behaviour is observed for $\tilde{C}_{mean}$
(see the green lines of Fig. \ref{fig:fig-3}).  Hence, the mean of
coherence for the optimal UQSD reduces to $C_{mean}^{opt} \equiv
2|\gamma|\big(1 - |\gamma|\big)$ [Eq. (\ref{eq:mean value})] and
$\tilde{C}_{mean}^{opt} \equiv 2\sqrt{|\gamma|\big(1 -
|\gamma|\big)}$ [Eq. (\ref{eq:mean2})] because it has the highest
probability of success at $|\gamma| = |\alpha_1|^2 = |\alpha_2|^2$.
Note that $C_{mean}^{opt}$ (or $\tilde{C}_{mean}^{opt}$) is the value of mean coherence for the optimal UQSD protocol.
Thus, a discrimination strategy or protocol will be an optimal UQSD if the value of mean coherence equals $C_{mean}^{opt}$ (or $\tilde{C}_{mean}^{opt}$).

If $p_1 \neq p_2$, we can measure the coherence for each result $i$ and compare it to
$2\sqrt{\frac{p_2}{p_1}}|\gamma|(1 -
\sqrt{\frac{p_2}{p_1}}|\gamma|)$ for $i=1$ and
$2\sqrt{\frac{p_1}{p_2}}|\gamma|(1 -
\sqrt{\frac{p_1}{p_2}}|\gamma|)$ for $i=2$ to determine the
optimality of UQSD (see Appendix B). 
If the measured values of coherence equal the values above, one can implement the optimal strategy to discriminate the given quantum states.

The above discussion can also be extended to general $d$ described
in Eq. (\ref{eq:joint unitary transformation}). As in Eq.
(\ref{eq:upper bound success probability}), with $\gamma_{ij} =
\iinner{\phi_i}{\phi_j}$, the upper bound for the success
probability of the UQSD is given by $P_s \leq 1 - \frac{1}{d-1}
\sum_{i, j\neq i} \sqrt{p_ip_j}|\gamma_{ij}|$.
This inequality can be saturated provided $p_1|\alpha_1|^2 =
p_2|\alpha_2|^2 = \cdots = p_d|\alpha_d|^2$, because $|\alpha_i|^2 =
\sqrt{\frac{p_j}{p_i}}|\gamma_{ij}|$ for all $i\neq j$,
therefore,
\bea
P_s &=& 1 - \sum_{i} p_i|\alpha_i|^2 = 1- \sum_i \big[\frac{1}{d-1}\sum_{j\neq i}p_i\sqrt{\frac{p_j}{p_i}}|\gamma_{ij}|\big] \nonumber \\
&=& 1- \frac{1}{d-1} \sum_{i, j\neq i}\sqrt{p_ip_j}|\gamma_{ij}|.
\eea Thus, for any $i \in \{1,2,...,d\}$, for example, $i=1$, we
have $d~p_1|\alpha_1|^2 = \frac{1}{d-1} \sum_{i, j\neq i}
\sqrt{p_ip_j}|\gamma_{ij}| \equiv B$.
Therefore, the mean of coherence is
\begin{eqnarray}\label{eq:mean value-3}
C_{mean} &=& 2B\Big(1 - \frac{B}{d^2} \sum_i \frac{1}{p_i} \Big).
\end{eqnarray}

However, since this upper bound of success probability is not always achievable, it cannot in general be regarded as an optimal success probability. Likewise, we cannot be certain that the mean of coherence in Eq. (\ref{eq:mean value-3}) is for an optimal discrimination.
It is only possible to estimate how similar or close our UQSD is to the optimal UQSD by comparing the computed mean value with that in Eq. (\ref{eq:mean value-3}). However, when the quantum states $\{\ket{\phi_i}\}$ satisfy the following two conditions, we can obtain the optimal result.

Condition 1. $\frac{|\gamma_{ij}||\gamma_{ik}|}{|\gamma_{jk}|} = \frac{|\gamma_{il}||\gamma_{im}|}{|\gamma_{lm}|}$ for unequal $i,j,k,l,m$.
This makes it possible for all $\ket{\varphi_i}$ to be equal in Eq. (\ref{eq:joint unitary transformation}), i.e.,
$U_{SA}\ket{\phi_i} \ket{0}_A = \sqrt{1-|\alpha_i|^2}\ket{\varphi} \ket{i}_A + \alpha_i\ket{\varphi} \ket{0}_A,$
where $\alpha_i^\ast\alpha_j = \iinner{\phi_i}{\phi_j}$.
Then $\frac{|\gamma_{1j}||\gamma_{1k}|}{|\gamma_{jk}|} = |\alpha_1|^2$ for any $j\neq k.$
%

Condition 2. $p_i|\gamma_{ki}|^2 = p_j|\gamma_{kj}|^2$ for unequal $i,j,k$.
This allows us to design a strategy which satisfies $p_1|\alpha_1|^2 = p_2|\alpha_2|^2 = \cdots = p_d|\alpha_d|^2$.

Therefore, when the above two conditions are satisfied, we can
verify that the UQSD is optimal by comparing the mean value of the
measured coherence with Eq. (\ref{eq:mean value-3}).

\section{The protocol with noise}

Next, we try to understand how the UQSD protocol using quantum coherence is affected when the input state is subject to noise. This noise can be modelled in a variety of ways, and depends on the actual implementation of the relevant devices. In the literature, arguably the most popular theoretical model of noise is admixture with white noise.
But as in our protocol of UQSD there is already a bias in the input state, it is plausible that the environmental noise will thereby be biased as well.
We restrict ourselves, in the noisy scenario, to the case where there are
two inputs to the distinguishing device, and they are respectively
$\ket{0}$ and $\ket{+}$, where \(\langle + | 0 \rangle = 1/\sqrt{2}\).
We assume the noise model where the density matrices corresponding to the states $\ket{0},\ket{+}$ become $\rho_0 = p\out{0}{0} + \frac{1-p}{2}\tilde{\id}_{2}$ and $\rho_+ = p\out{+}{+} + \frac{1-p}{2}\tilde{\id}_{2}$,
where $\tilde{\id}_{2} = \out{0}{0} + \out{+}{+}$. Calculating the final state after the unitary transformation, we see that no entanglement or discord is generated. And, for various values of the noise parameter $(1-p)$, we have calculated the value of quantum coherence. Please refer to the Appendix C for the detailed analysis. We also explicitly show that the reliability of the distinguishing protocol decreases from $1$ to $\frac{1+p}{2}$ in the presence of noise, where $(1-p)$ is the strength of the noise.

\section{Conclusion}
Identifying resources for quantum state
discrimination is of fundamental importance. Use of quantum
correlations as a resource for the same has been studied
extensively. In this paper, we have investigated the role of
quantum coherence in unambiguously discriminating nonorthogonal but
linearly independent pure quantum states, assisted by an auxiliary
system.
We provide a relationship between the success probability of the
discriminating strategy and the mean coherence generated on the
auxiliary system for several important coherence measures. The
degree of the generated coherence depends on the nonorthogonality
between the input quantum states.
We can effectively use the mean of coherence to improve the
efficiency of the strategy for each individual result of the
performed measurement. Finally, we compute the coherence that is
generated when an optimal unambiguous discrimination strategy is
implemented in some situations. In these cases, we can use the mean
of coherence to determine whether the discrimination strategy is
optimal or not. In particular, for unambiguous discrimination
between two pure qubit states, we show that the receiver can obtain the
optimal strategy by controlling the mean coherence to the maximum or
minimum value without feedback from the sender. Our result will open
up new investigations in the use of coherence in quantum state
discrimination.

\section{Appendix}
\label{appendix}

\subsection{The joint unitary operators for $d\geq 2$ quantum states}\label{joint unitary transformation}

We construct a unitary operator $U_{SA}$
in the UQSD strategy that discriminates between two quantum states $\ket{\phi_1
}$ and $\ket{\phi_2}$.
Let $\iinner{\phi_1}{\phi_2} = \gamma$ and $\{\ket{i}_A\}_{i=0}^2$ be
an orthonormal basis of the auxiliary system $A$. We assume that
the system $S$ is 2 dimensional. Take a vector $\ket{\phi_1^+}\in S$
such that $\iinner{\phi_1}{\phi_1^+}=0$ and $\ket{\phi_2} =
\gamma\ket{\phi_1} + \sqrt{1-|\gamma|^2}\ket{\phi_1^+}$. Then
$\{\ket{\phi_1}\ket{i}_A, \ket{\phi_1^+}\ket{i}_A\}_i$ is an
orthonormal basis of the whole system $SA$. Let $0<|\alpha|\leq 1$,
$\ket{\upsilon_{\alpha,1}}_A = \sqrt{1-|\alpha|^2}\ket{1}_A +
\alpha\ket{0}_A$ and $\ket{\upsilon_{\alpha, 2}}_A =
\frac{\gamma(1-|\alpha|^2)}{\alpha^\ast\sqrt{1-|\gamma|^2}}\ket{0}_A
- \frac{\gamma\sqrt{1-|\alpha|^2}}{\sqrt{1-|\gamma|^2}}\ket{1}_A +
\frac{\sqrt{|\alpha|^2 -
|\gamma|^2}}{|\alpha|\sqrt{1-|\gamma|^2}}\ket{2}_A$. Then it is easy
to see that $\iinner{\upsilon_{\alpha,
1}}{\upsilon_{\alpha,2}}_A = 0$. Moreover, let us take a unit vector
$\ket{\upsilon_{\alpha,0}}_A\in A$ such that
$\{\ket{\upsilon_{\alpha,j}}_A\}_{j=0}^2$ is an orthonormal basis of
the auxiliary system $A$, and take two unit vectors $\ket{\varphi},
\ket{\varphi^+}$ in $S$ such that $\iinner{\varphi}{\varphi^+} =
0$. Let $U_{\gamma,\alpha}$ be a transform from the orthonormal basis
$\{\ket{\phi_1}\ket{i}_A, \ket{\phi_1^+}\ket{i}_A\}_i$ to the
orthonormal basis $\{\ket{\varphi}\ket{\upsilon_{\alpha,j}}_A,
\ket{\varphi^+}\ket{\upsilon_{\alpha,j}}_A\}_j$ that satisfies
$U_{\gamma,\alpha}\ket{\phi_1}\ket{0}_A = \ket{\varphi}\ket{\upsilon_{\alpha,1}}_A \ \textmd{and} \
U_{\gamma,\alpha}\ket{\phi_1^+}\ket{0}_A = \ket{\varphi}\ket{\upsilon_{\alpha,2}}_A.$
Then
$U_{\gamma,\alpha}$ is a unitary transformation on the system $SA$
such that:
\bea
U_{\gamma,\alpha}\ket{\phi_1}\ket{0}_A &=& \sqrt{1-|\alpha|^2}\ket{\varphi}\ket{1}_A + \alpha\ket{\varphi}\ket{0}_A,\\
U_{\gamma,\alpha}\ket{\phi_2}\ket{0}_A &=&
\sqrt{1-\frac{|\gamma|^2}{|\alpha|^2}}\ket{\varphi}\ket{2}_A +
\frac{\gamma}{\alpha^\ast}\ket{\varphi}\ket{0}_A.
\eea

Next, let us find out about the case of $d\geq 3$. For $d$ linearly independent quantum states $\ket{\phi_i}$ with $\iinner{\phi_i}{\phi_j} = \gamma_{ij}$, the unitary operator $U_{SA}$ of (\ref{eq:joint unitary transformation}) can be achieved by expanding and repeating similar tasks as above, but it must include more complex process. We first find the following orthonormal basis $\{\ket{\phi_i'}\}$ of the system $S$ sequentially from the states $\ket{\phi_i}$:
\be\label{eq:8}
\ket{\phi_1'} = \ket{\phi_1} \quad \textmd{and} \quad
\ket{\phi_i'} = \frac{\ket{\phi_i} - \sum_{j=1}^{i-1} \gamma_{ji}'\ket{\phi_j'}}{\sqrt{1-\sum_{j=1}^{i-1}|\gamma_{ji}'|^2}}
\ee
with $\iinner{\phi_j'}{\phi_i} = \gamma_{ji}'$ for $2\leq i \leq d$. For this we also know that $\ket{\phi_i}$ can be represented as a combination of $\{\ket{\phi_i'}\}$, \emph{i.e.},
\be\label{eq:9}
\ket{\phi_i} = \sum_{j=1}^{i-1}\gamma_{ji}'\ket{\phi_j'} + \sqrt{1-\sum_{j=1}^{i-1}|\gamma_{ji}'|^2}\ket{\phi_i'}
\ee
for $2\leq i \leq d$. Our aim here is to find states $\ket{\varphi_i'}$ and $\ket{\upsilon_{\alpha_i}}_A$ with $\iinner{\upsilon_{\alpha_i}}{\upsilon_{\alpha_j}}_A= 0$ for $i\neq j$ that satisfy the Eq. (\ref{eq:joint unitary transformation}) when $U_{SA}\ket{\phi_i'}\ket{0}_A=\ket{\varphi_i'}\ket{\upsilon_{\alpha_i}}_A$. These states can be found sequentially, starting with $\ket{\varphi_1'} = \ket{\varphi_1}$ and $\ket{\upsilon_{\alpha_1}} = \sqrt{1-|\alpha_1|^2}\ket{1}+ \alpha_1\ket{0}$. Of course, $\alpha_i^\ast\alpha_j\iinner{\varphi_i}{\varphi_j}=\iinner{\phi_i}{\phi_j}$ means that the inner products between states before and after the unitary $U_{SA}$ are preserved, and therefore the existence of $U_{SA}$ satisfying the Eq. (\ref{eq:joint unitary transformation}) is guaranteed.

In addition, if $\alpha_i$ are satisfied with $\alpha_i^\ast\alpha_j = \iinner{\phi_i}{\phi_j}$ for $i\neq j$, we can find $U_{SA}$ in a simpler way. Here we also use the orthonormal basis $\{\ket{\phi_i'}\}$ of the system $S$ in (\ref{eq:8}), and find quantum states $\ket{\upsilon_{\alpha_i}}$ with $\iinner{\upsilon_{\alpha_i}}{\upsilon_{\alpha_j}}=0$ for $(i\neq j)$ satisfying that
$$\sum_{j=1}^{i-1}\gamma_{ji}'\ket{\upsilon_{\alpha_j}} + \sqrt{1-\sum_{j=1}^{i-1}|\gamma_{ji}'|^2}\ket{\upsilon_{\alpha_i}} = \sqrt{1-|\alpha_i|^2}\ket{i}+ \alpha_i\ket{0}.$$
In the above equation, the part before the equal sign is the same as the form for $\ket{\phi_i'}$ in (\ref{eq:9}) and thereby we can find the states $\ket{\upsilon_{\alpha_i}}$ in sequence, starting with $\ket{\upsilon_{\alpha_1}} = \sqrt{1-|\alpha_1|^2}\ket{1}+ \alpha_1\ket{0}$.
Then the joint unitary operator $U_{SA}$ that result in $U_{SA}\ket{\phi_i'}\ket{0}_A = \ket{\varphi}\ket{\upsilon_{\alpha_i}}_A$ for any state $\ket{\varphi}$ satisfy Eq. (\ref{eq:joint unitary transformation}).

In addition, if $\alpha_i$ are satisfied with $\alpha_i^\ast\alpha_j = \iinner{\phi_i}{\phi_j}$ for $i\neq j$, we can find $U_{SA}$ in a simpler way. Here we also use the orthonormal basis $\{\ket{\phi_i'}\}$ of the system $S$ in (\ref{eq:8}), and find quantum states $\ket{\upsilon_{\alpha_i}}$ with $\iinner{\upsilon_{\alpha_i}}{\upsilon_{\alpha_j}}=0$ for $(i\neq j)$ satisfying that
$$\sum_{j=1}^{i-1}\gamma_{ji}'\ket{\upsilon_{\alpha_j}} + \sqrt{1-\sum_{j=1}^{i-1}|\gamma_{ji}'|^2}\ket{\upsilon_{\alpha_i}} = \sqrt{1-|\alpha_i|^2}\ket{i}+ \alpha_i\ket{0}.$$
In the above equation, the part before the equal sign is the same as the form for $\ket{\phi_i'}$ in (\ref{eq:9}) and thereby we can find the states $\ket{\upsilon_{\alpha_i}}$ in sequence, starting with $\ket{\upsilon_{\alpha_1}} = \sqrt{1-|\alpha_1|^2}\ket{1}+ \alpha_1\ket{0}$.
Then the joint unitary operator $U_{SA}$ that result in $U_{SA}\ket{\phi_i'}\ket{0}_A = \ket{\varphi}\ket{\upsilon_{\alpha_i}}_A$ for any state $\ket{\varphi}$ satisfy Eq. (\ref{eq:joint unitary transformation}).

\subsection{Relation of $P_s$, $C_{mean}$ and $\tilde{C}_{mean}$ for two quantum states}

The success probability to discriminate between two quantum states
$\ket{\phi_1 }$ and $\ket{\phi_2}$ is
\begin{eqnarray}
\label{eq:success probability} P_s &=& 1-
\tr{\id\ot\ket{0}_A\bra{0}\rho} = \sum_{i=1}^{2} p_i(1-|\alpha_i|^2).
\end{eqnarray}

Note that for the $U_{SA}$ in the above section, we have
$|\alpha_1|^2|\alpha_2|^2= |\gamma|^2$. Denoting $x = |\alpha_1|^2$,
$P_s(x) = p_1(1 - x) + p_2 \big(1 - \frac{|\gamma|^2}{x} \big).$
For the optimal success probability, we require $P_s'(x) = -p_1 +
p_2\frac{|\gamma|^2}{x^2}=0$. This yields $p_1|\alpha_1|^2 =
p_2|\alpha_2|^2$. That is, if
$|\alpha_1|^2=\sqrt{\frac{p_2}{p_1}}|\gamma|$, then we can
distinguish $\ket{\phi_1 }$ and $\ket{\phi_2}$ with the optimal
success probability \bea P_s^{opt} = 1 - 2p_1|\alpha_1|^2 = 1 -
2\sqrt{p_1p_2}|\gamma|. \eea Since $p_1p_2|\gamma|^2 = p_1^2
|\alpha_1|^4 = p_2^2 |\alpha_2|^4$, the mean of coherence, $C_{mean}
= 2\sum_{i=1}^d p_i |\alpha_i|^2 \big(1 - |\alpha_i|^2\big)$,
for the optimal UQSD is $$C^{opt}_{mean} = 2|\gamma| (2\sqrt{p_1p_2} - |\gamma|).$$

When $p_1 = p_2 = \frac{1}{2}$, we have
$$C_{mean}(x) = \big[x(1 - x) + \frac{|\gamma|^2}{x} \big(1 - \frac{|\gamma|^2}{x} \big) \big],$$
and its first-order derivative with respect to x
$$C_{mean}'(x) = -\frac{1}{x^3}(x - |\gamma|)(x+|\gamma|)(2x^2
-x + 2|\gamma|^2).$$
Thus $C_{mean}'(x) = 0$ has three roots:
$x_1=\gamma$, $x_2=\frac{1+\sqrt{1-16|\gamma|}}{4}$, and
$x_3=\frac{1+\sqrt{1-16|\gamma|}}{4}$, where $|\gamma| \leq \frac{1}{4}$.

Moreover, from the second-order derivative of $C_{mean}(x)$ with respect to x
$$C_{mean}''(x) = - 2 + 2\frac{|\gamma|^2}{x^3} - 6\frac{|\gamma|^4}{x^4}$$
it follows that
\[C_{mean}''(x)\Big|_{|\gamma| = x} \left\{
  \begin{array}{lr}
   \leq 0 ~~~~~~~~\text{when}~~|\gamma|\geq\frac{1}{4},\\
    > 0 ~~~~~~~~\text{when}~~|\gamma| < \frac{1}{4}.
  \end{array}
\right.
\]
Therefore, there is only one extreme point at $x = |\gamma| =
|\alpha_1|^2$ when $|\gamma|\geq \frac{1}{4}$.

Furthermore, for the second-type mean of coherence $\tilde{C}_{mean}(x)$ when $p_1 = p_2 = \frac{1}{2}$, we have
$$\tilde{C}_{mean}(x) = \sqrt{x-x^2} + |\gamma|\sqrt{\frac{1}{x}-
\frac{|\gamma|^2}{x^2}}$$ and $$\tilde{C}_{mean}'(x) =
\frac{1}{2}\Big[\frac{1-2x}{\sqrt{x-x^2}} -
\frac{|\gamma|(x-2|\gamma|^2)}{x^2\sqrt{x-|\gamma|^2}}\Big].$$
Thus, $\tilde{C}_{mean}$ also has an extreme value at $x = |\gamma|
= |\alpha_1|^2$.

\subsection{UQSD using quantum coherence in presence of noise}
\label{appendix:noise}

Suppose that the states to be distinguished using UQSD are $\ket{0}$ and $\ket{+}$.
We consider the noise model where the density matrices corresponding to the states $\ket{0},\ket{+}$, become
\begin{eqnarray}
\rho_0 &=& p\out{0}{0} + \frac{1-p}{2}\tilde{\id}_{2},  \mathrm{ \hspace{0.3cm}  and}\\
\rho_+ &=& p\out{+}{+} + \frac{1-p}{2}\tilde{\id}_{2},
\end{eqnarray}
where $\tilde{\id}_{2} = \out{0}{0} + \out{+}{+}$. It is possible to calculate the degree of quantum coherence of a quantum state if its spectral decomposition is known. Therefore, we first obtain the spectral decomposition of the above quantum states for any noise $1-p ~(0\leq p\leq1)$ as follows:
\begin{eqnarray*}
\rho_0 &=& q\out{\psi_+}{\psi_+} + (1-q)\out{\psi_-}{\psi_-}\\
\rho_+ &=& q\out{\phi_+}{\phi_+} + (1-q)\out{\phi_-}{\phi_-},
\end{eqnarray*}
where
\begin{eqnarray*}
\ket{\psi_+} &=& a_0\ket{0} + \sqrt{1-a_0^2}\ket{1},\ \ \ket{\psi_-} = \sqrt{1-a_0^2}\ket{0} - a_0\ket{1},\\
\ket{\phi_+} &=& a_+\ket{0} + \sqrt{1-a_+^2}\ket{1},  \ket{\phi_-} = \sqrt{1-a_+^2}\ket{0} - a_+\ket{1}
\end{eqnarray*}
with $a_0^2 = \frac{1}{2} + \frac{\sqrt{2}(1+p)}{4\sqrt{1+p^2}}, a_+^2 = \frac{1}{2} + \frac{\sqrt{2}(1-p)}{4\sqrt{1+p^2}}$ and $q = \frac{1}{2} + \frac{\sqrt{2}\sqrt{1+p^2}}{4}$.  If $U_{SA}$ is a joint unitary transformation on the system $SA$
such that:
\bea
|\Phi_{0}\rangle=U_{SA}\ket{0}\ket{0}_A &=& \sqrt{1-\alpha^2}\ket{\varphi_{0}}\ket{1}_A + \alpha\ket{\varphi_{0}}\ket{0}_A,\\
|\Phi_{+}\rangle=U_{SA}\ket{+}\ket{0}_A &=&
\sqrt{1-\frac{\gamma^2}{\alpha^2}}\ket{\varphi_{+}}\ket{2}_A +
\frac{\gamma}{\alpha}\ket{\varphi_{+}}\ket{0}_A,
\eea
where $\gamma = \iinner{0}{+} = \frac{\sqrt{2}}{2}$,
then the action of the unitary operator $U_{SA}$ on the states $\rho_{0}$ and $\rho_{+}$ is
\begin{eqnarray}
\label{unitary1}
&&U_{SA}(\rho_0 \ot \ket{0}_A\bra{0})U_{SA}^\dagger \nonumber \\
&&=\frac{(1+p)}{2}|\Phi_{0}\rangle\langle \Phi_{0}| + \frac{1-p}{2}|\Phi_{+}\rangle\langle \Phi_{+}|,
\end{eqnarray}

\begin{eqnarray}
\label{unitary2}
&&U_{SA}(\rho_+ \ot \ket{0}_A\bra{0})U_{SA}^\dagger \nonumber \\
&&=\frac{(1+p)}{2}|\Phi_{+}\rangle\langle \Phi_{+}| + \frac{1-p}{2}|\Phi_{0}\rangle\langle \Phi_{0}|.
\end{eqnarray}

The unitary transformations in Eq. (\ref{unitary1}) and Eq. (\ref{unitary2}) can be rewritten as
\begin{eqnarray}
&&U_{SA}(\rho_i \ot \ket{0}_A\bra{0})U_{SA}^\dagger =\out{\varphi_{i}}{\varphi_{i}} \ot \rho^A_{i} \nonumber \\
&& = \out{\varphi_{i}}{\varphi_{i}}\ot \Big\{q\ket{\Phi_{i,+}}_A\bra{\Phi_{i,+}}+(1-q)\ket{\Phi_{i,-}}_A\bra{\Phi_{i,-}}\Big\}, \nonumber \\
\end{eqnarray}
where
\begin{eqnarray*}
\ket{\Phi_{i,+}}_A &=& \Big[\big\{\alpha(a_i - \sqrt{1-a_i^2})+\frac{1}{\alpha}\sqrt{1-a_i^2}\big\}\ket{0}\\
&& + \sqrt{1-\alpha^2}(a_i-\sqrt{1-a_i^2})\ket{1}\\
&& + \sqrt{2-\frac{1}{\alpha^2}}\sqrt{1-a_i^2}\ket{2}\Big]_A
\end{eqnarray*}
and
\begin{eqnarray*}
\ket{\Phi_{i,-}}_A &=& \Big[\big\{\alpha(a_i + \sqrt{1-a_i^2})-\frac{a_i}{\alpha}\big\}\ket{0}\\
&& + \sqrt{1-\alpha^2}(a_i+\sqrt{1-a_i^2})\ket{1}\\
&& - \sqrt{2-\frac{1}{\alpha^2}}a_i\ket{2}\Big]_A,
\end{eqnarray*}
for $i = 0,+$. We can see that no quantum entanglement or discord is generated after the unitary transformation for any noise $1-p$.

Next, we calculate quantum coherence with respect to measurement $\Pi = \{\ket{0}_A\bra{0}, \ket{1}_A\bra{1}, \ket{2}_A\bra{2}\}$ on system A,
\begin{eqnarray*}
C_I(\rho_i^A) &=& \sum_{j=0}^2 \Big[q\big\{(+)_{i,j}^2-(+)_{i,j}^4\big\} + (1-q)\big\{(-)_{i,j}^2-(-)_{i,j}^4\big\}\\
&& - 2\sqrt{q(1-q)}(+)_{i,j}^2(-)_{i,j}^2\Big],
\end{eqnarray*}
where $(+)_{i,j} = \iinner{\Phi_{i,+}}{j}_A, (-)_{i,j} = \iinner{\Phi_{i,-}}{j}_A$ for $i=0,+$ and $j=0,1,2$.
In addition, with some tedious calculation, we can predict the value of noise $1-p$ from the measured values of quantum coherence when $\alpha$ is fixed.

For example, when $p=1$ (the case of no noise), we have
\begin{eqnarray*}
C_I(\rho_0^A) = 2\alpha^2(1-\alpha^2), \quad C_I(\rho_+^A) = \frac{1}{\alpha^2}(1-\frac{1}{2\alpha^2}).
\end{eqnarray*}
This is the measured value of quantum coherence in the absence of noise.
If $\alpha^2 = \gamma =\frac{\sqrt{2}}{2}$, then we have
\begin{eqnarray*}
p=1 \ \quad &:& \quad C_I(\rho_0^A) = C_I(\rho_+^A) \approx 0.414,\\
p=0.5 \quad &:& \quad C_I(\rho_0^A) = C_I(\rho_+^A) \approx 0.287,\\
p=0.2 \quad &:& \quad C_I(\rho_0^A) = C_I(\rho_+^A) \approx 0.271,\\
p=0 \ \quad &:& \quad C_I(\rho_0^A) = C_I(\rho_+^A) \approx 0.269.
\end{eqnarray*}

Now, if we happen to know that the probability of the measurement outcome is $|0\rangle$, we can calculate the probability with which the input state $|0\rangle$ was sent. We call this probability {\it reliability} when $|0\rangle$ clicks in measurement $M$, and denote it by $R_{0}$ (see Ref. \cite{Sedlak}). The expression for $R_{0}$ is given by
\begin{widetext}
\begin{eqnarray}
\label{Bayes_rule}
&&R_{0}= \textrm{Pr}\big(|0\rangle \textrm{was sent} \big| \textrm{outcome of $M$ is} |0\rangle \big) \nonumber \\
&&\quad~ = \frac{\textrm{Pr}\big( |0\rangle \textrm{was sent}) \times \textrm{Pr}(\textrm{outcome of $M$ is} |0\rangle \big| |0\rangle \textrm{was sent} \big)}{\textrm{Pr}\big( |0\rangle \textrm{was sent}) \times \textrm{Pr}(\textrm{outcome of $M$ is} |0\rangle \big| |0\rangle \textrm{was sent} \big) + \textrm{Pr} \big( |+\rangle \textrm{was sent}) \times \textrm{Pr}(\textrm{outcome of $M$ is} |0\rangle \big| |+\rangle \textrm{was sent} \big)}, \nonumber \\
\end{eqnarray}
\end{widetext}
where $\mathrm{Pr}(\cdot)$ denotes the probability of the event in the argument, and the Bayes rule \cite{fellar, gupta} is used in the second line of Eq. (\ref{Bayes_rule}). Assuming that the states $|0\rangle$ and $|1\rangle$ were chosen with equal probabilities, we get
\begin{eqnarray}
\label{R_{0}}
R_{0} = \frac{\frac{1}{2}\times p_{0}}{\frac{1}{2} \times p_{0} + \frac{1}{2} \times p_{+}} = \frac{p_{0}}{p_{0}+p_{+}},
\end{eqnarray}
where
\begin{eqnarray}
p_{0} = {}_{A}\langle 0| \textrm{Tr}_{S} \bigg(\frac{1+p}{2} |\Phi_{0}\rangle\langle \Phi_{0}| + \frac{1-p}{2}|\Phi_{+}\rangle\langle \Phi_{+}| \bigg) |0\rangle_{A}, \nonumber \\   p_{+} = {}_{A}\langle 0| \textrm{Tr}_{S} \bigg(\frac{1+p}{2} |\Phi_{+}\rangle\langle \Phi_{+}| + \frac{1-p}{2}|\Phi_{0}\rangle\langle \Phi_{0}| \bigg) |0\rangle_{A}. \nonumber
\end{eqnarray}
Since
\begin{eqnarray}
\textrm{Tr}_{S}|\Phi_{0}\rangle\langle \Phi_{0}\rangle = \mathbb{P}[(1-|\alpha_{0}|^{2})|0\rangle + \alpha_{0}|2\rangle] \nonumber \\
\textrm{and    }\textrm{Tr}_{S}|\Phi_{+}\rangle\langle \Phi_{+}\rangle = \mathbb{P}[(1-|\alpha_{+}|^{2})|1\rangle + \alpha_{+}|2\rangle], \nonumber
\end{eqnarray}
where $\mathbb{P}(\cdot)$ denotes the projector of the vector in the argument, we can write
\begin{eqnarray}
p_{0} = \frac{1+p}{2} (1-|\alpha_{0}|^{2}), \nonumber \\
p_{+} = \frac{1-p}{2} (1-|\alpha_{0}|^{2}) .
\end{eqnarray}
Substituting these in Eq. (\ref{R_{0}}), we get
\begin{equation}
\label{R_{0}again}
R_{0} = \frac{1+p}{2}.
\end{equation}
Performing a similar analysis for the reliability $R_{+}$, when $|+\rangle$ clicks in $M$, we obtain
\begin{equation}
R_{+} = R_{0} = \frac{1+p}{2},
\end{equation}
which can be called the ``reliability of the entire distinguishing process''.
Hence, we can say that the reliability of the distinguishing process decreases from $1$ to $\frac{1+p}{2}$, when noise acts on the system, where $1-p$    $(0\leq p \leq 1)$ is strength of the noise for the noise model under consideration. Note that the reliability is lower bounded by $1/2$.

\begin{acknowledgments}
This project is supported by the National Natural Science Foundation of China (Grants No. 12050410232, 12031004  and 61877054). The research of SD is supported in part by the INFOSYS scholarship for senior students. We are grateful to E. Andersson for suggesting an alternative protocol for unambiguous quantum state discrimination discussed in this paper.
\end{acknowledgments}


\begin{thebibliography}{99}
\bibitem{note2} Irrespective of the protocol used to perform the UQSD between the nonorthogonal but linearly independent set of states, the average coherence in the auxiliary system, summed over the set of states, cannot vanish, as that would require the auxiliary states to be orthogonal in the basis of the measurement, which in turn would make the set of nonorthogonal states exactly distinguishable, violating the no-cloning theorem.
\end{thebibliography}

%

\end{document}